\documentclass[twocolumn,tighten,twocolappendix]{aastex631}
\usepackage{CJK}
\usepackage{amsbsy}
\usepackage{multirow}
\usepackage{bm}
\usepackage{ulem}
\usepackage{graphicx}     
\usepackage{chngcntr}     

\usepackage{epsfig}
\usepackage{color}
\usepackage{graphicx}
\usepackage{longtable}
\usepackage{hyperref}
\usepackage{amsmath}

\usepackage[mathlines]{lineno}

\def\e{\,{\boldsymbol{e_1}}}

\received{XXX}
\revised{YYY}
\accepted{ZZZ}
\published{HHH}

\counterwithout{figure}{section}

\begin{document}
\begin{CJK*}{UTF8}{gbsn}

\title{Observational Evidence for the Kinematic Memory of Cosmic Filaments from Satellite Orbital Orientations}

\correspondingauthor{Peng Wang}
\email{pwang@shao.ac.cn}

\author[0000-0003-2504-3835]{Peng Wang (王鹏)}
\affil{Shanghai Astronomical Observatory, Chinese Academy of Sciences, Nandan Road 80, Shanghai 200030, People's Republic of China.}

\author[0000-0002-8350-6423]{Wei Wang (王伟)}
\affil{Shanghai Astronomical Observatory, Chinese Academy of Sciences, Nandan Road 80, Shanghai 200030, People's Republic of China.}

\author[0000-0002-5762-7571]{Wenting Wang (王文婷)}
\affil{Key Laboratory for Particle Astrophysics and Cosmology (MOE)/Shanghai Key Laboratory for Particle Physics and Cosmology, Shanghai 200240, People's Republic of China.}
\affil{State Key Laboratory of Dark Matter Physics, School of Physics and Astronomy, Shanghai Jiao Tong University, Shanghai 200240, People's Republic of China.}

\author[0009-0005-9342-9125]{Min Bao (鲍敏)}
\affil{School of Physics and Technology, Nanjing Normal University, Nanjing 210023, People's Republic of China.}

\author[0000-0002-5458-4254]{Xi Kang (康熙)}
\affil{Institute for Astronomy, School of Physics, Zhejiang University, Hangzhou 310027, People's Republic of China.}
\affil{Center for Cosmology and Computational Astrophysics, Zhejiang University, Hangzhou 310027, People's Republic of China.}

\author[0000-0003-4936-8247]{Hong Guo (郭宏)}
\affil{Shanghai Astronomical Observatory, Chinese Academy of Sciences, Nandan Road 80, Shanghai 200030, People's Republic of China.}

\author[0000-0003-1967-4091]{Youcai Zhang (张友财)}
\affil{Shanghai Astronomical Observatory, Chinese Academy of Sciences, Nandan Road 80, Shanghai 200030, People's Republic of China.}

\author[0009-0001-7527-4116]{Xiao-xiao Tang (唐潇潇)}
\affil{Shanghai Astronomical Observatory, Chinese Academy of Sciences, Nandan Road 80, Shanghai 200030, People's Republic of China.}

\begin{abstract}
We present an observational study of the kinematic coherence between satellite orbital planes and the cosmic web. Using the SDSS DR12 galaxy sample combined with the Bisous filament catalogue, we investigate whether the orbital motion of satellites preserves the memory of filamentary accretion. For each satellite system, we define a projected orbital-normal vector using galaxy sky positions and line-of-sight velocity offsets. By measuring the angle $\theta$ between this vector and the local projected filament direction, we detect a distinctive preferred orientation: satellite orbital planes tend to contain or lie parallel to the filament axis. This signal deviates from the isotropic expectation at a high significance level of $12.8\sigma$. The strength of this kinematic connection depend strongly on environment and host properties. The preference for orbital planes to track the filament direction is most pronounced for groups in close distance to the filament spine and for more massive hosts. Conversely, at intermediate distances from the filament and at large group-centric radii, the signal reverses, indicating a tendency for orbital planes to be oriented perpendicular to the filament. Our findings provide direct observational evidence for the two-phase model of filamentary accretion, where a transition from initial perpendicular collapse toward the filament spine to subsequent parallel streamwise infall into dark matter haloes governs the orientation of satellite orbital angular momentum and galaxy spin. The observed transition may further trace the characteristic radial scale of filaments, offering a dynamical perspective on the internal structure and assembly of the cosmic filament.
\end{abstract}



\keywords{
    \href{http://astrothesaurus.org/uat/584}{Galaxy clusters (584)};
    \href{http://astrothesaurus.org/uat/902}{Large-scale structure of the universe (902)};
    \href{http://astrothesaurus.org/uat/1882}{Astrostatistics (1882)}}


\section{Introduction} 
\label{sec:intro}

The anisotropic growth of structure is one of the most distinctive predictions of the cosmic web paradigm \citep{1970A&A.....5...84Z, 1996Natur.380..603B, 2018MNRAS.473.1195L}. In the standard $\Lambda$CDM framework, galaxies, groups, and clusters form through hierarchical accretion along an interconnected network of sheets, filaments, and nodes \citep{1984ApJ...286...38W, 2005Natur.435..629S}. Among these structures, filaments play a particularly important role because they channel matter and angular momentum from large scales into dark matter haloes \citep{2011MNRAS.418.2493P,2014MNRAS.443.1274L, 2015ApJ...813....6K}. As a result, the orientation of galaxies and haloes with respect to the cosmic web has long been regarded as a fossil record of structure formation \citep[e.g.,][]{1969ApJ...155..393P, 1996MNRAS.281...84V, 1984ApJ...286...38W}.

A large body of work has investigated the alignment between galaxy or halo spin and the surrounding filamentary environment. Both simulations \citep[e.g.,][]{2007MNRAS.381...41H,2007MNRAS.375..489H, 2009ApJ...706..747Z,2012MNRAS.427.3320C,2018MNRAS.481.4753C, 2018MNRAS.481..414G,2019MNRAS.487.1607G,2021MNRAS.503.2280G} and observations \citep[e.g.,][]{2013ApJ...779..160Z, 2013MNRAS.428.1827T, 2016MNRAS.457..695P,2025ApJ...987L..30W,2025ApJ...995L...8M} have shown that halo and galaxy spins exhibit non-random correlations with filaments, although the sign and strength of the alignment depend on halo mass \citep{2013ApJ...762...72T}, galaxy type \citep{2013MNRAS.428.1827T,2025ApJ...983..122R,2025A&A...703A..11M}, redshift \citep{2018ApJ...866..138W}, and the scale \citep{2014MNRAS.440L..46A} on which the filament is defined. Low-mass haloes are often found to have spin vectors preferentially parallel to filaments, while massive haloes tend to show a perpendicular configuration, a transition \citep[e.g.,][]{2014MNRAS.444.1453D,2018ApJ...866..138W,2020MNRAS.491.2864W} commonly interpreted as the consequence of anisotropic mergers \citep{2014MNRAS.445L..46W} and nonlinear tidal torques \citep{2013ApJ...766L..15L}. These studies have established that the cosmic web leaves an imprint on the internal angular momentum of haloes and galaxies.

However, the angular momentum of a galaxy system is not only encoded in the spin of its central galaxy or host halo. Satellite galaxies orbiting within groups and clusters also carry orbital angular momentum \citep[e.g.,][]{2005ApJ...624..505Z,2011MNRAS.413.3013L,2018MNRAS.476.1796S}, which directly traces the kinematic memory of their accretion history. If satellites are preferentially accreted along filaments \citep[e.g.,][]{2014MNRAS.443.1274L, 2015ApJ...807...37S,2022MNRAS.516.4576D}, their orbital motions should not be randomly oriented with respect to the local filament direction. Instead, the ensemble orbital plane of satellites is expected to retain a coherent relation to the filament through which they were supplied. In this picture, the filament acts as an accretion channel: satellites fall in preferentially along the filament axis, and their projected orbital plane should statistically contain the filament direction. Equivalently, the normal vector of the satellite orbital plane is expected to be preferentially perpendicular to the filament.

Despite its physical simplicity, this prediction has received much less direct observational attention than spin--filament alignments. Most previous observational studies have focused on the orientation of central galaxies\cite[][]{2013ApJ...779..160Z}, satellite galaxy distribution\citep[][]{2020ApJ...900..129W}. The projected orbital angular momentum of satellite systems offers a complementary and more dynamical probe. Unlike galaxy shapes/spins, it uses both the projected spatial distribution of satellites and their line-of-sight velocities, and therefore provides direct access to the internal orbital structure of galaxy groups and clusters. This makes it possible to test whether the present-day kinematics of satellite systems still preserve the anisotropic infall pattern imposed by the cosmic web.

Recent observational studies have shown that satellite galaxies can be used to trace the projected angular momentum of galaxy groups \citep{2025ApJ...995L...9T} and to investigate its connection with the large-scale environment. \cite{2025ApJ...983L...3R} and \cite{2025JCAP...10..095W} explored the alignment between the group spin inferred from satellite galaxies and the surrounding cosmic filament. These results indicate that the dynamical properties of satellite systems can retain information about the anisotropic assembly history of their host groups. This connection can be understood within the anisotropic-collapse framework originally introduced by \cite{1970A&A.....5...84Z}. Numerical studies have revealed that the flow of matter within and around filaments exhibits a characteristic transition \citep{2014MNRAS.441.2923C}. This physical picture was subsequently formulated as a two-phase accretion model by \cite{2015ApJ...813....6K} and \cite{2018MNRAS.473.1562W} to explain the mass-dependent halo spin--filament alignment. In this scenario, matter first collapses toward filaments preferentially from directions perpendicular to the filament spine and is subsequently transported along filaments toward dark matter haloes. Such an anisotropic assembly history may leave a characteristic imprint on the orbital motions of satellite galaxies.

In this Letter, we provide a direct observational test of the kinematic coherence between the projected orbital angular momentum of satellite galaxies and the projected direction of their host filaments. By combining projected positions with line-of-sight velocities, we aim to isolate the dynamical imprint of anisotropic accretion and determine the extent to which satellite systems retain a memory of their host filaments.



\section{Data and Methodology}
\label{sec:method}

\subsection{Galaxy, Group and Filament Catalogs}

Our analysis is based on the spectroscopic galaxy sample from the Sloan Digital Sky Survey Data Release 12 \citep[SDSS DR12;][]{2015ApJS..219...12A}, using the group catalog constructed by \citet{2017A&A...602A.100T}. 
We apply a Petrosian $r$-band magnitude cut of $m_r \leq 17.77$\,mag, resulting in a parent sample of 584,449 galaxies in the redshift range $0.015 \leq z \leq 0.2$ ($z_{\rm med} = 0.09$).

Galaxy groups are identified using a Friends-of-Friends algorithm \citep{1985ApJ...295..368B, 2009MNRAS.399..497D} with redshift-dependent linking lengths \citep{2014A&A...566A...1T}. 
The group catalog contains 88,662 systems. 
Within each system, the brightest galaxy is defined as the central, and the remaining members are treated as satellites. The group masses $M_{200}$ and the richness $N_{\rm gal}$ are taken from the catalog \citep{2017A&A...602A.100T}, where $M_{200}$ is estimated from the dispersion velocity of the member galaxies.

The cosmic filament catalog was constructed from the full SDSS DR12 galaxy sample, including both central and satellite galaxies, using the Bisous stochastic marked point process \citep{2014MNRAS.438.3465T,2016A&C....16...17T}. As a robustness check, we repeat our main analysis using filaments identified with the alternative \textsc{DisPerSE} method; the corresponding results are presented in Appendix~\ref{app:sec1}.
For each filament segment, the catalog provides a three-dimensional unit vector $\hat{\boldsymbol{e}}_{\rm fil}^{\rm 3D} = (e_x, e_y, e_z)$ that describes the local filament orientation. For each system, the filament direction is assigned by associating the central galaxy with its nearest filament segment via a three-dimensional \textsc{cKDTree} search. 
To enable a direct comparison with the projected orbital-normal vector, the filament direction is projected onto the sky plane at the position of the central galaxy. 
Denoting the line-of-sight unit vector as $\hat{\boldsymbol{r}}_{\rm cen}$, the projected filament vector is given by
\begin{equation}
\boldsymbol{e}_{\rm fil}^{\rm proj}
=
\hat{\boldsymbol{e}}_{\rm fil}^{\rm 3D}
-
\left(
\hat{\boldsymbol{e}}_{\rm fil}^{\rm 3D}
\cdot
\hat{\boldsymbol{r}}_{\rm cen}
\right)
\hat{\boldsymbol{r}}_{\rm cen}, \nonumber
\end{equation}
and normalized to obtain the unit vector $\hat{\boldsymbol{e}}_{\rm fil}$. The perpendicular distance to the filament spine, $d_{\rm fil}$, is also recorded for each system and used in subsequent analysis.

\subsection{Projected Satellite Orbital Angular Momentum}

We define the projected orbital angular momentum of satellite galaxies using quantities that are directly accessible to observations. Our procedure for computing the projected angular momentum closely follows that employed in previous works \citep{2025ApJ...983L...3R, 2025ApJ...992L..17W}. In each galaxy group, we take the central galaxy as the origin of the projected system-centric coordinate system and use its redshift to define the systemic redshift of the system. For each satellite galaxy $i$，its projected position relative to the central galaxy is given by $\boldsymbol{r}_i$ on the sky plane. The line-of-sight velocity offset is computed as
\begin{equation}
\Delta v_i = c \, \frac{z_i - z_{\rm cen}}{1+z_{\rm cen}},
\end{equation}
where $z_i$ and $z_{\rm cen}$ are the redshifts of the satellite and central galaxy, respectively. $c$ is the speed of light. The projected orbital angular momentum of the satellite system is then estimated as
\begin{equation}
\boldsymbol{L}_{\rm orb}^{\rm obs}
=
\sum_{i=1}^{N_{\rm sat}}
\boldsymbol{r}_i \times \Delta v_i,
\end{equation}
This vector is perpendicular, in projection, to the characteristic projected orbital plane traced by the satellite population. We therefore define the unit projected orbital-normal vector as
\begin{equation}
\hat{\boldsymbol{n}}_{\rm orb}
=
\frac{\boldsymbol{L}_{\rm orb}^{\rm obs}}
{\left| \boldsymbol{L}_{\rm orb}^{\rm obs} \right|}.
\end{equation}
Although we acknowledge that the measurement of the projected orbital angular momentum is naturally sensitive to the number of available satellite galaxies, to maintain a consistent and fair comparison, we define our primary sample to consist of groups with at least one satellite galaxy ($N_{\rm gal} \geq 1$) and the distance from the filament of $d_{\rm fil} \leq 10 \, \mathrm{Mpc}$. 
Applying these selection criteria results in a fiducial sample of 76, 959 systems. In the subsequent analysis, we further investigate how the statistical signal is influenced by both the satellite count and the group--filament separation.

\subsection{Signal Characterization and Validation}

We quantify the relative orientation between the orbital-normal vector of the satellite system and the direction of the filament using the angle $\theta$, defined as
\begin{equation}
\theta = \arccos \left( \left| \hat{\boldsymbol{n}}_{\rm orb} \cdot \hat{\boldsymbol{e}}_{\rm fil} \right| \right),
\end{equation}
where $0^\circ \leq \theta \leq 90^\circ$. The absolute value is employed because both the orbital-normal vector and the filament direction are axial quantities in projection; reversing either vector does not alter the physical configuration.

To assess the statistical significance of any deviation from isotropy in the measured angles, we compare the observed distribution of $\theta$ with the isotropic expectation. For two random unoriented axes in a two-dimensional projected plane, $\theta$ is uniformly distributed over the interval $0^\circ$--$90^\circ$. The expectated mean angle is $\left< \theta \right>_{\rm iso} = 45^\circ$.

In order to assess the significance of the signal, we construct a randomized null hypothesis by destroying the physical correspondence between satellite system orbital-normal vectors and filament directions while preserving the sample selection and group properties. In practice, the projected filament directions are randomly reassigned among systems, and the resulting distribution of $\theta$ is recomputed. This procedure is repeated 1000 times to estimate the mean and scatter of the null distribution in each angular bin. The observed probability density is then compared with the randomized expectation.

We quantify the deviation from isotropy using both the mean measured angle and the full distribution of $\theta$. 
The significance of the mean signal is estimated as
\begin{equation}
S =
\frac{
\left< \theta \right>
-
\left< \theta \right>_{\rm iso}
}{
\sigma_{\left< \theta \right>_{\rm iso}}
},
\end{equation}
where $\sigma_{\left< \theta \right>_{\rm iso}}$ is the uncertainty on the observed mean angle, estimated from bootstrap resampling of the group sample. In addition, we perform a Kolmogorov--Smirnov test between the observed $\theta$ distribution and the isotropic expectation. The KS probability provides a non-parametric measure of whether the full observed distribution can be drawn from the null hypothesis.

\subsection{Physical Interpretation of the Orbital Geometry}

A statistical comparison between the observed distribution and this isotropic baseline provides a direct physical interpretation of the satellite accretion geometry. 

Specifically, a measured mean angle of $\langle \theta \rangle < 45^\circ$ indicates that $\hat{\boldsymbol{n}}_{\rm orb}$ is preferentially parallel to $\hat{\boldsymbol{e}}_{\rm fil}$, implying that satellite orbital planes are oriented perpendicular to the filament axis and that satellites are preferentially accreted from directions approximately perpendicular to the filament. Conversely, a measured value of $\langle \theta \rangle > 45^\circ$ indicates a preferentially perpendicular configuration between $\hat{\boldsymbol{n}}_{\rm orb}$ and $\hat{\boldsymbol{e}}_{\rm fil}$; in this case, the satellite orbital planes are aligned with the filament, suggesting that satellite accretion occurs preferentially along the filament axis. 

Rather than asserting a definitive interpretation at this stage, we treat systematic departures of $\langle \theta \rangle$ from $45^\circ$ as observable signatures to discriminate between competing accretion scenarios. Such behavior which was so called two-phase accretion model was initially reported by \cite{2015ApJ...813....6K} in simulations and more fully explored by \cite{2018MNRAS.473.1562W} as a possible explanation for the spin-flip from low-mass to high-mass in the alignment between dark matter halo spin and the cosmic filament. By measuring $\langle \theta \rangle$ as a function of host and filament properties, we aim to provide an observational test of whether these predicted transitions in accretion geometry can be identified in our sample.

\section{Result}
\label{sec:results}

\subsection{Signal of fiducial sample}
\begin{figure}[t]
\centering
\plotone{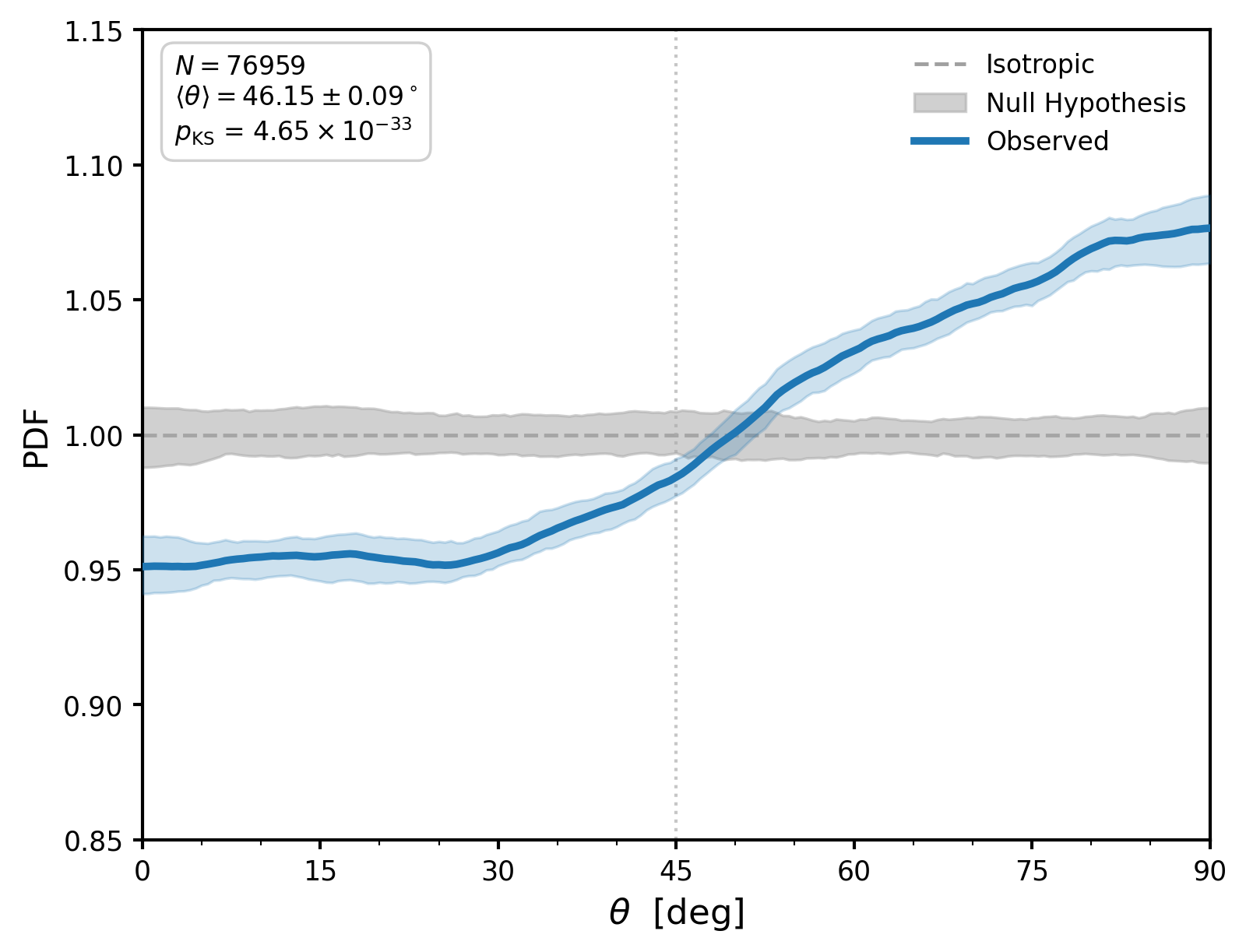}
\caption{
Probability density distribution (PDF) of the alignment angle $\theta$ between the projected satellite orbital-normal vector $\hat{n}_{\rm orb}$ and the local filament direction $\hat{e}_{\rm fil}$. 
The blue solid line shows the kernel density estimation (KDE) of the observed $\theta$ distribution for the sample of 76,959 satellite systems. We use an Epanechnikov kernel with a bandwidth of $10^\circ$ and apply boundary reflection at $\theta=0^\circ$ and $90^\circ$. The blue shaded region represents the $1\sigma$ uncertainty estimated from 1000 bootstrap resamplings.
The horizontal grey dashed line  denotes the analytical expectation for an isotropic distribution. 
The grey shaded band indicates the $1\sigma$ null hypothesis range, generated from 1,000 random realizations of the filament orientations. 
The vertical dotted line marks the $\theta = 45^\circ$ isotropic midpoint. 
Statistics including the mean alignment angle $\langle\theta\rangle \pm \sigma_{\langle\theta\rangle}$ and the Kolmogorov-Smirnov (KS) test $p$-value against a uniform distribution are shown in the legend.
}
\label{fig:pdf_alignment}
\end{figure}

The statistical measurement of satellite orbital orientations relative to their host filaments reveals a distinctive kinematic structure that reflects the large-scale environment. Figure~\ref{fig:pdf_alignment} presents the probability density function (PDF) of the angle $\theta$ between the projected satellite orbital-normal vector $\hat{n}_{\rm orb}$ and the local filament direction $\hat{e}_{\rm fil}$ for our fiducial sample of $N=76{,}959$ satellite systems.

For an isotropic distribution of orientations, $\theta$ is expected to be uniformly distributed over $[0^\circ, 90^\circ]$. We construct a randomized null hypothesis by shuffling the filament orientations across groups, which reproduces the isotropic expectation (horizontal dashed line) within the $1\sigma$ uncertainty band (grey shaded region). As shown in Figure~\ref{fig:pdf_alignment}, the observed PDF (blue solid line) deviates markedly and systematically from this isotropic benchmark. The distribution exhibits a clear monotonic increase toward larger angles, crossing the isotropic midpoint at $\theta = 45^\circ$ with a prominent excess at $\theta > 45^\circ$ and a corresponding deficit at $\theta < 45^\circ$. This departure is highly significant, as the observed signal remains well outside the $1\sigma$ null hypothesis range across the majority of the angular domain.

Quantitatively, we measure a mean angle of
$\langle \theta \rangle = 46.15^\circ \pm 0.09^\circ$
which exceeds the isotropic expectation of $45^\circ$ at a significance of $12.78\sigma$. A Kolmogorov-Smirnov test yields $p_{\rm KS} = 4.65 \times 10^{-33}$, robustly rejecting the null hypothesis of random orientations.

Since the orbital-normal vector $\hat{n}_{\rm orb}$ is by definition perpendicular to the orbital plane, the observed excess at large $\theta$ (approaching $90^\circ$) signifies that the normal vectors tend to be perpendicular to the filament spine. Geometrically, this implies that satellite galaxies preferentially orbit within planes that are aligned with their host filaments. This detection provides compelling observational evidence that the satellite population maintains a coherent kinematic structure rather than being randomly oriented. The preference for satellites to orbit within planes aligned with the filament direction is consistent with previous results regarding anisotropic accretion and the alignment of satellite systems with the large-scale structure \citep[e.g.,][]{2014MNRAS.443.1274L,2015ApJ...807...37S,2015ApJ...813....6K,2017MNRAS.472.4099K,2018MNRAS.473.1562W,2022MNRAS.516.4576D}. These results suggest that cosmic filaments serve as primary channels for mass assembly, funneling satellites into groups with significant directional persistence.

\begin{figure*}[ht]
\centering
\plottwo{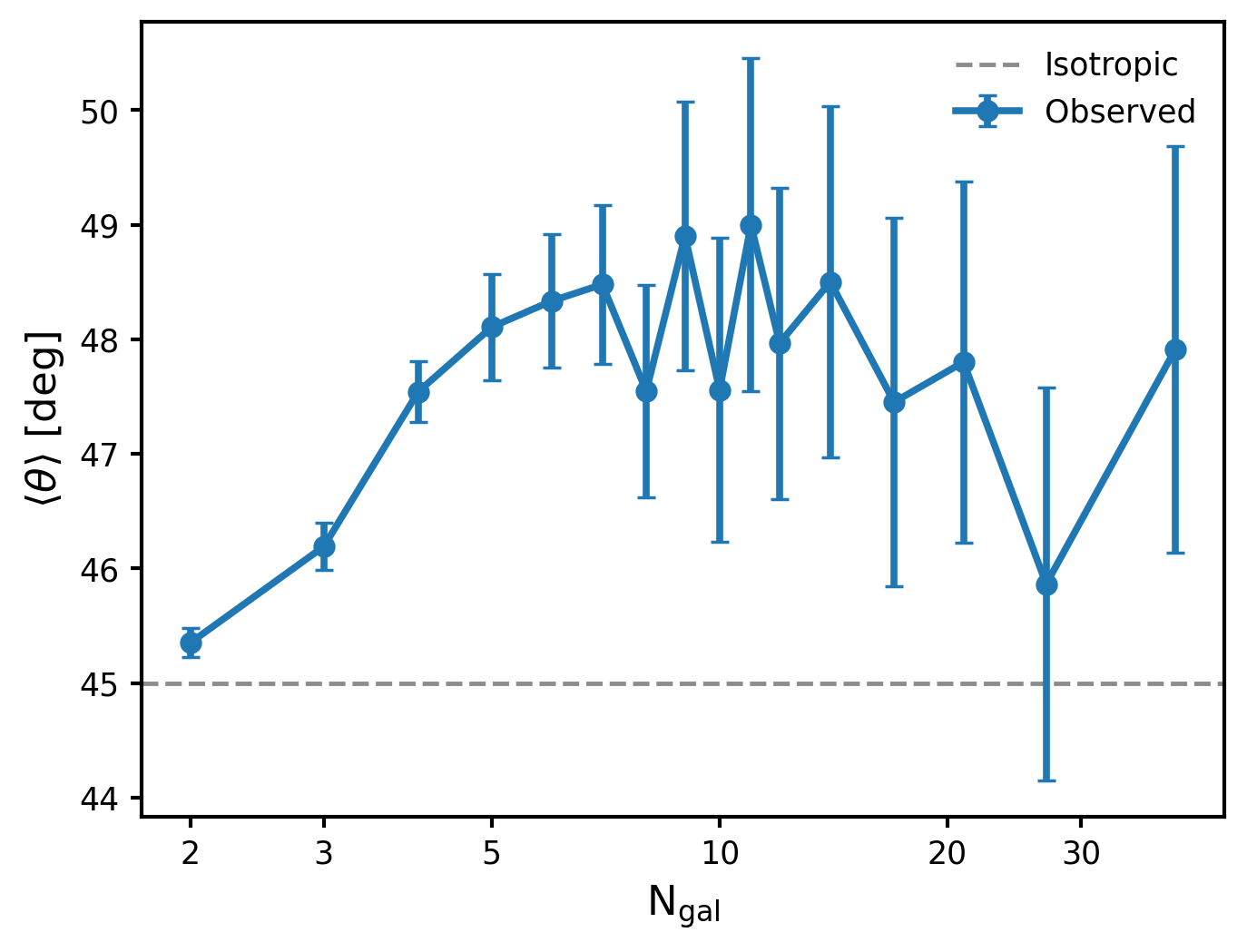}{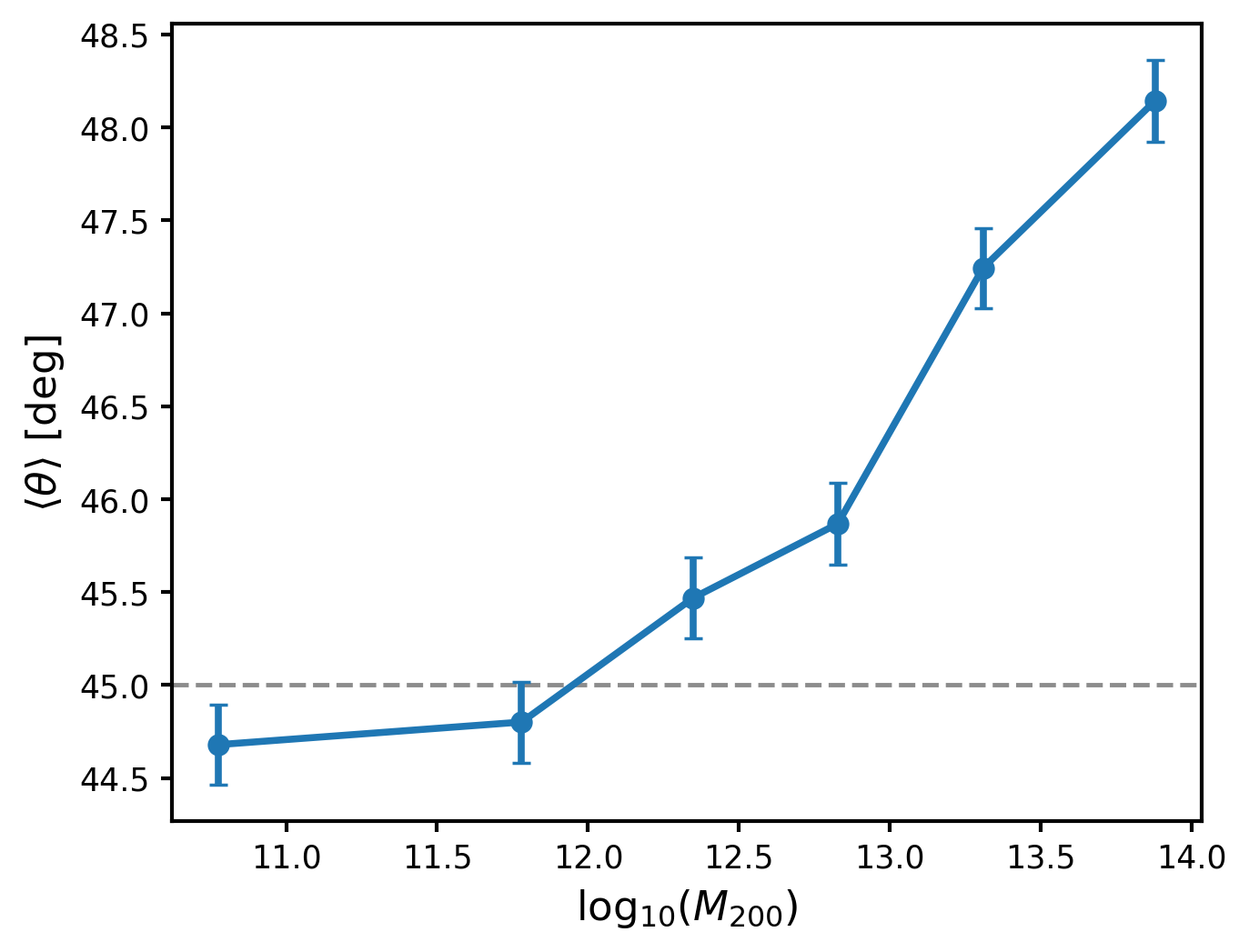}
\plottwo{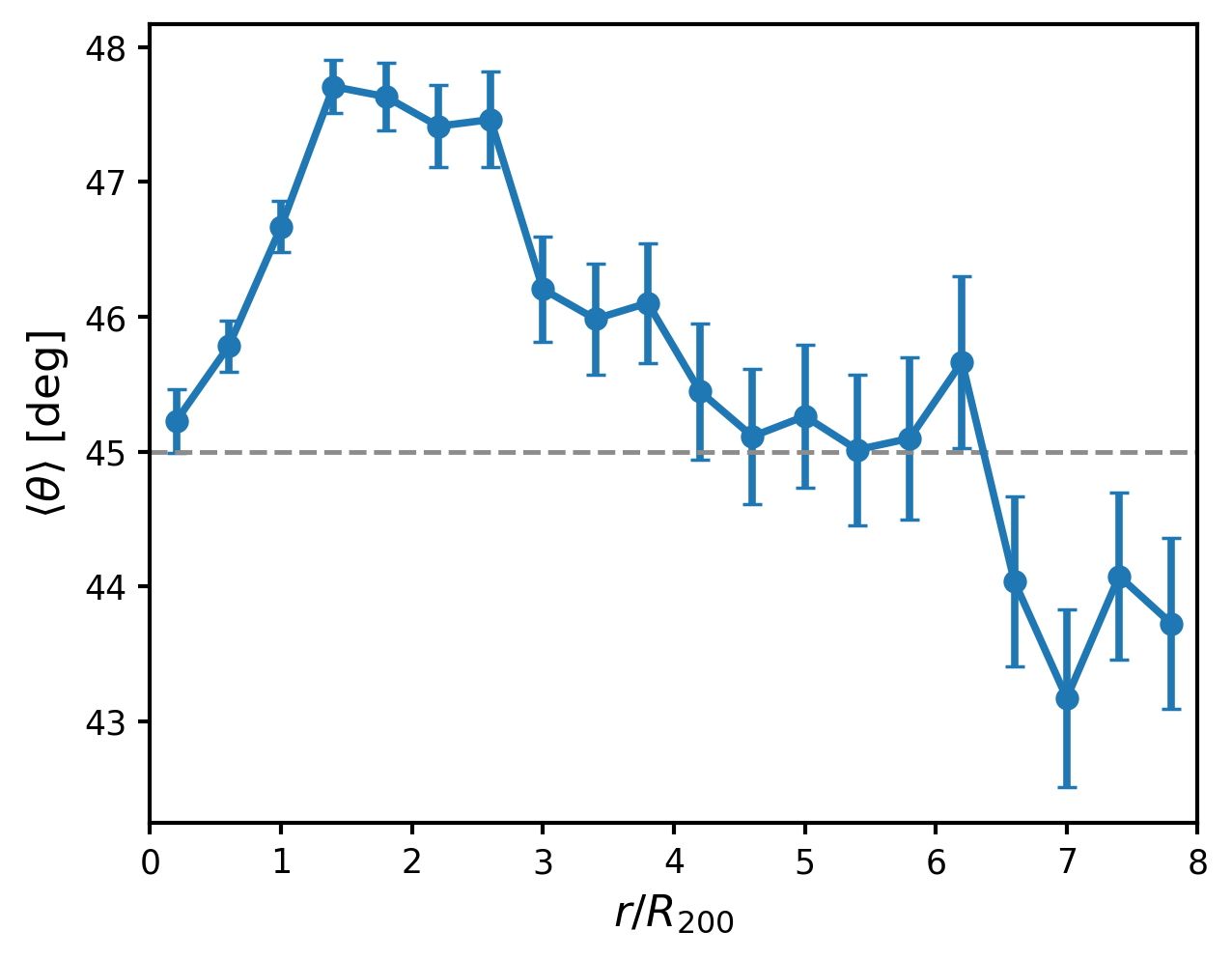}{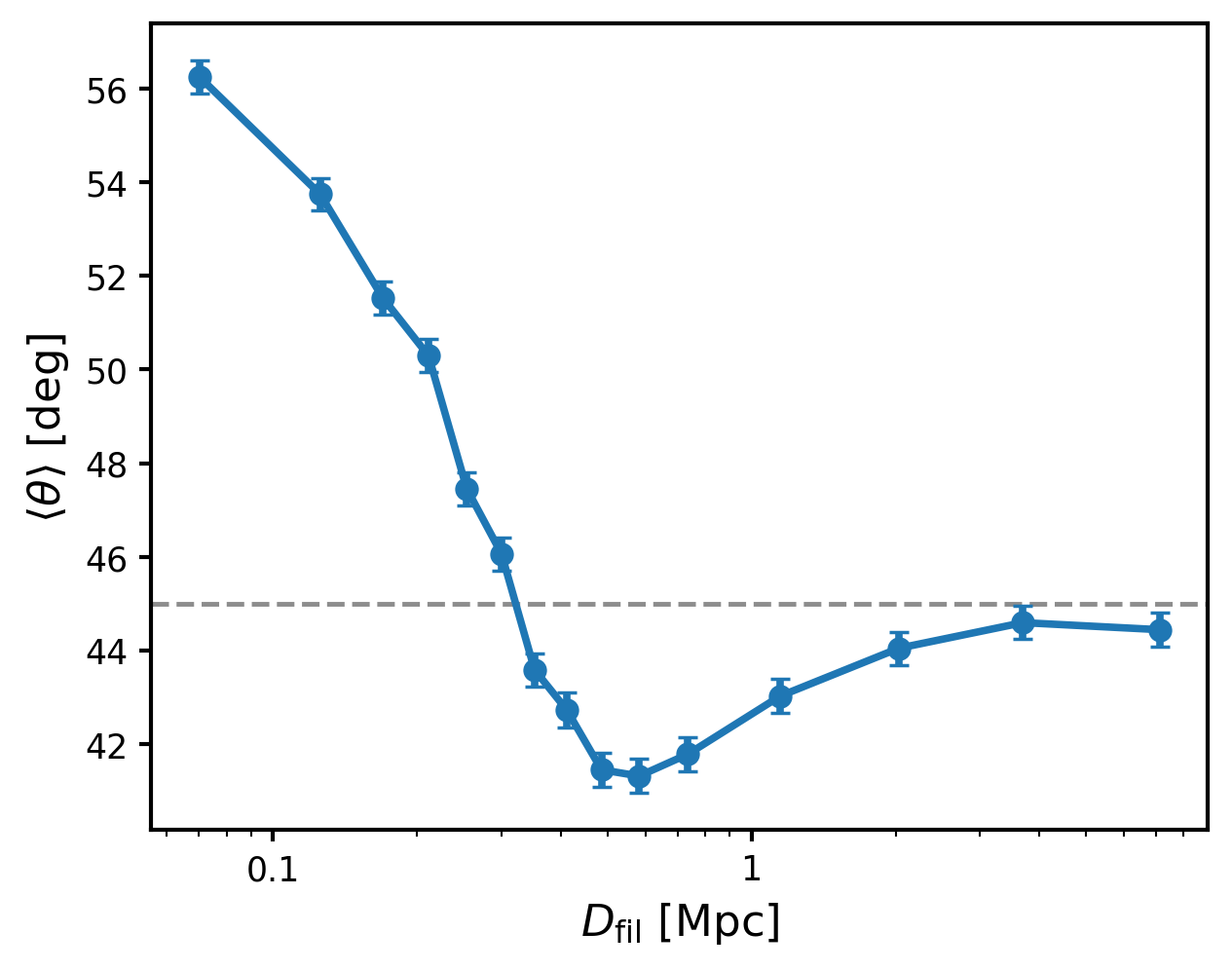}
\caption{The dependence of the angle $\theta$, between the satellite orbital-normal vector $\hat{\boldsymbol{n}}_{\rm orb}$ and the projected filament direction $\hat{\boldsymbol{e}}_{\rm fil}$, on 
group richness $N_{\rm gal}$ \textbf{(Upper-left panel)}, 
host halo mass $\log_{10}(M_{200})$\textbf{(Upper-right panel)}, 
the normalized radial distance $r/R_{200}$  \textbf{(Bottom-left panel)} and 
the projected distance to the nearest filament $D_{\rm fil}$ \textbf{(Bottom-right panel)}. 
The grey dashed line in all panels denotes the isotropic expectation ($\langle\theta\rangle = 45^\circ$). Blue points represent the binned mean values of the angle $\theta$, with error bars indicating the $1\sigma$ uncertainty on the mean.}
\label{fig:signal_dependence}
\end{figure*}

\subsection{Signal dependence on group and environmental properties}
While the signal in Figure~\ref{fig:pdf_alignment} characterizes our fiducial sample as a whole, it remains to be seen how this kinematic coherence is influenced by specific properties of the groups and their surrounding environment. In this subsection, we investigate the variation of the mean angle $\langle\theta\rangle$ with respect to group richness, halo mass, and spatial scales to determine the key drivers of the observed signal. Figure~\ref{fig:signal_dependence} examines how the kinematic alignment signal depends on group and environmental properties. In all panels, the grey dashed line marks the isotropic expectation of $\langle\theta\rangle = 45^\circ$. Values of $\langle\theta\rangle > 45^\circ$ indicate that satellite orbital-normal vectors $\hat{\boldsymbol{n}}_{\rm orb}$ are preferentially perpendicular to the projected filament direction $\hat{\boldsymbol{e}}_{\rm fil}$, meaning the orbital planes tend to contain or be parallel to the filament. Conversely, values of $\langle\theta\rangle < 45^\circ$ indicate that orbital-normal vectors tend to align with the filament direction, implying that the orbital planes are preferentially perpendicular to the filament axis.

We first examine the dependence on group richness $N_{\rm gal}$, which directly impacts the robustness of the orbital-normal estimate. Since the reliability of $\hat{\bm{n}}_{\rm orb}$ is expected to improve with increasing $N_{\rm gal}$, this test is primarily a robustness check rather than a physical inference. The upper-left panel shows the mean angle as a function of group richness $N_{\rm gal}$. For the poorest groups ($N_{\rm gal} \approx 2$), the signal is only slightly above the isotropic expectation ($\langle\theta\rangle \approx 45.3^\circ$). As richness increases to $N_{\rm gal} \approx 5-15$, $\langle\theta\rangle$ rises and fluctuates between $47.5^\circ$ and $49^\circ$, suggesting that higher-richness groups (which are typically more massive) exhibit a stronger tendency for satellite orbits to align with the filament axis.

In the upper-right panel, $\langle\theta\rangle$ is shown to be a strong function of host halo mass $\log_{10}(M_{200})$. Systems with lower mass ($\log_{10}M_{200} \lesssim 11.8$) show signals consistent with or slightly below isotropy. However, for more massive haloes, the signal increases monotonically, reaching $\langle\theta\rangle \approx 48.2^\circ$ at the highest mass bin. This indicates that the alignment of orbital planes along filaments is enhanced in deeper gravitational potentials.

The dependence on the normalized projected radial distance $r/R_{200}$ from the system center is presented in the bottom-left panel. Here, we do not assign each group to a radial bin according to the mean $r/R_{200}$ of its satellite population. Instead, for each radial bin, we select the satellites of each group whose individual projected distances fall within that bin and recompute the projected orbital-normal vector using only this subset of satellites. Each group therefore contributes one alignment angle to a given radial bin and may contribute to multiple radial bins if its satellites occupy different radial ranges.
The signal exhibit a clear non-monotonic behavior: $\langle\theta\rangle$ increases from $\sim 45.2^\circ$ in the innermost region to a prominent peak of $\sim 47.7^\circ$ at intermediate radii ($1.5 \lesssim r/R_{200} \lesssim 2.5$). Moving further out, the signal gradually weakens, crossing the isotropic line at $r/R_{200} \approx 4-6$, and falling to less than $45^\circ$ at $r/R_{200}>6$, where orbits become preferentially perpendicular to the filament.

The most significant variation is observed with respect to the projected distance to the nearest filament $D_{\rm fil}$ (bottom-right panel). Starting from large separations, the signal stays close to the $45^\circ$ isotropic baseline, stabilizing around $44.5^\circ$; this indicates that at large distances, satellite motions are nearly random and the filamentary influence on their orbital geometry is relatively weak. As satellites approach the filamentary environment ($0.3 < D_{\rm fil} < 1 \text{ } h^{-1}\text{Mpc}$), $\langle\theta\rangle$ decreases, reaching a minimum of $\sim 41.5^\circ$ at $D_{\rm fil} \approx 0.5$--$0.7 \text{ } h^{-1}\text{Mpc}$. This preference for $\langle\theta\rangle < 45^\circ$ suggests a configuration where orbital planes are oriented perpendicular to the filament axis, reflecting the initial stage of accretion where satellites are pulled from the surrounding wall toward the filament from transverse directions. However, at the smallest scales ($D_{\rm fil} < 0.3 \text{ } h^{-1}\text{Mpc}$), the signal exhibits a dramatic ``flip'', crossing over $45^\circ$ and rising sharply to $\langle\theta\rangle \approx 56^\circ$. This transition to a strongly parallel configuration ($\langle\theta\rangle > 45^\circ$) implies that once satellites are deeply embedded within the filamentary spine, their orbital planes become aligned with the filament axis, consistent with a late-stage accretion scenario where satellites stream coherently along the filament toward the host group. Collectively, this scale-dependent ``swing'' in $\langle\theta\rangle$ illustrates the dynamic reorientation of satellite kinematic states as they transition from the isotropic cosmic web into the highly anisotropic filamentary environment.

The four panels show a consistent picture: the kinematic coherence between satellite orbital planes and the large-scale filament is strongest for satellites very close to filaments and in more massive / richer systems, while the dependence on radius within the group is non-monotonic, with a peak at intermediate radii followed by a decline toward the outskirts. These trends suggest that both filamentary accretion (proxied by \(D_{\rm fil}\)) and host properties (mass, richness) contribute to shaping satellite orbital alignments.

\begin{figure*}[t]
\centering
\plotthree{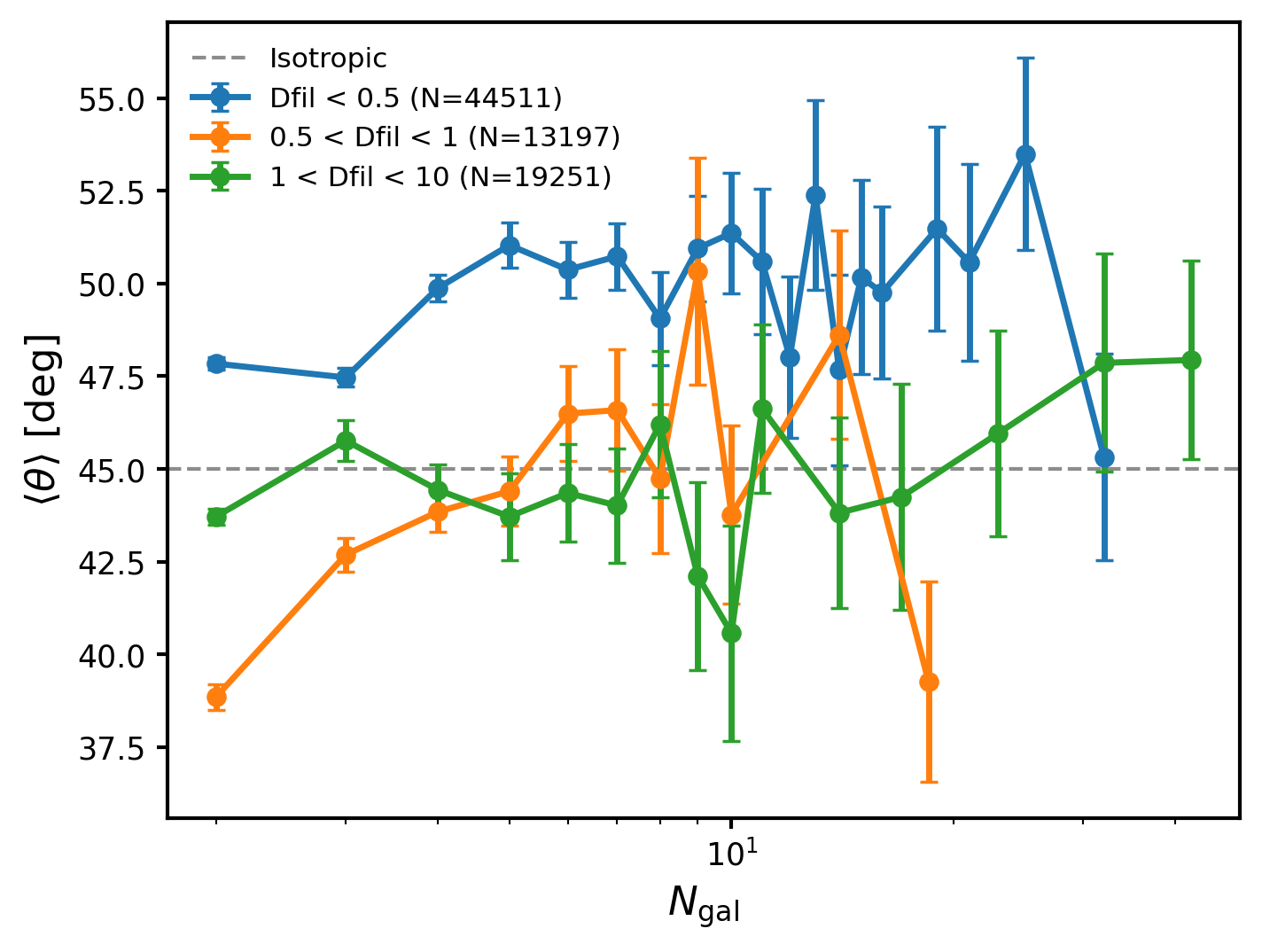}{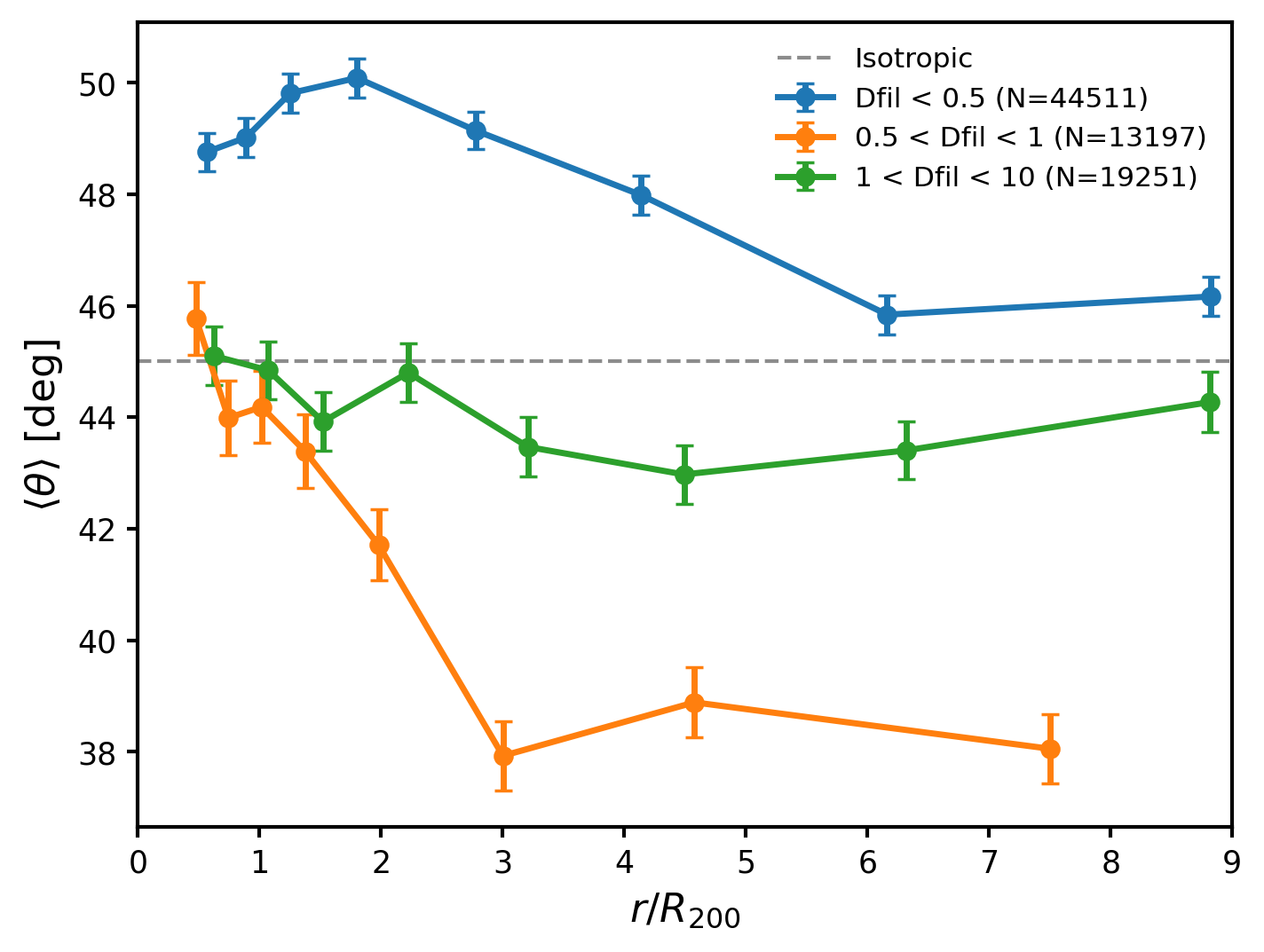}{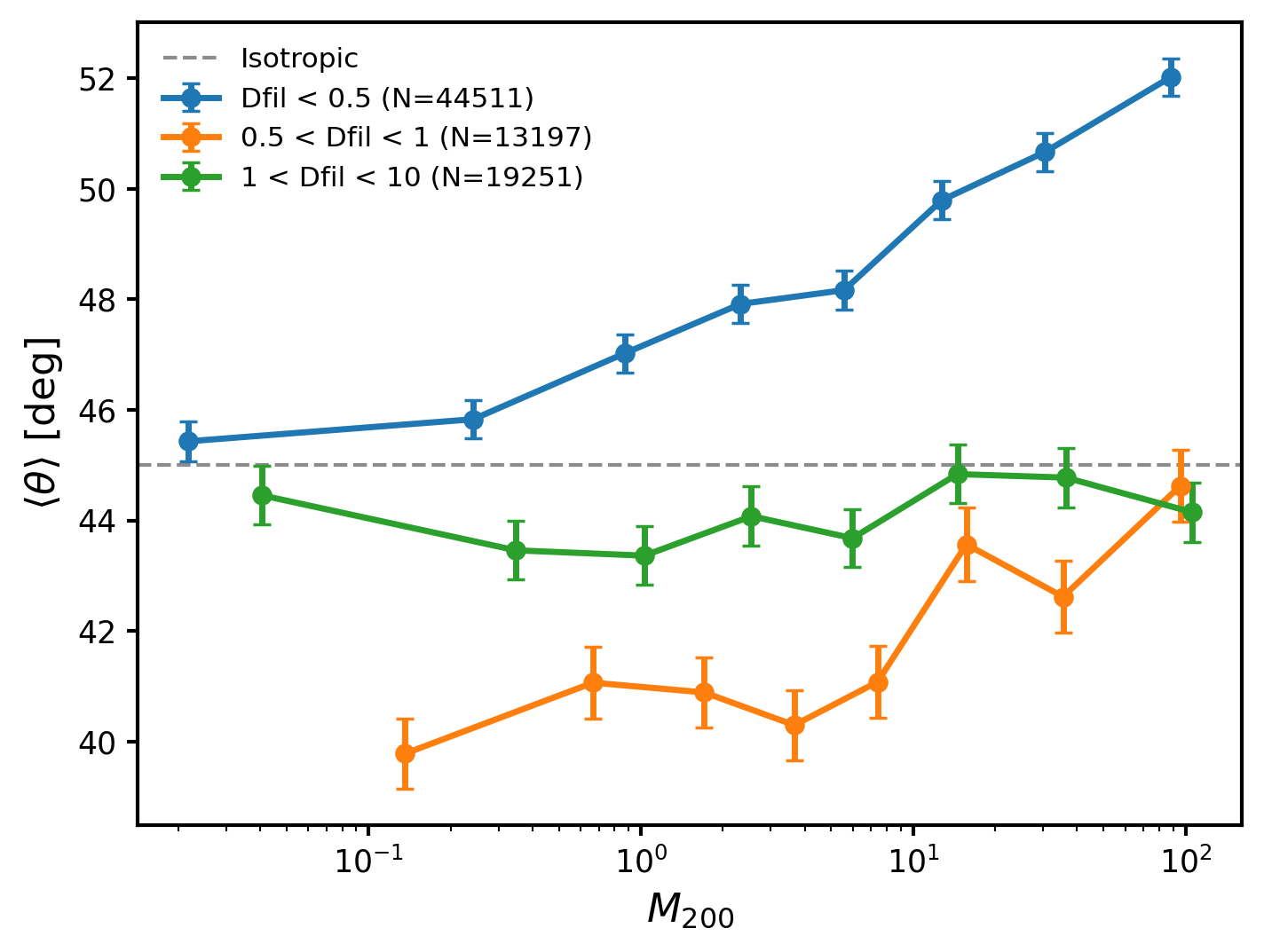}
\caption{Disentangling the primary drivers of the  signal. Similar as Figure~\ref{fig:signal_dependence}, but in each panel the sample is subdivided by projected distance to the nearest filament into three bins: $D_{\rm fil}<0.5\,h^{-1}\mathrm{Mpc}$ (blue), $0.5 \le D_{\rm fil} < 1.0\,h^{-1}\mathrm{Mpc}$ (orange), and $1.0 \le D_{\rm fil} < 10.0\,h^{-1}\mathrm{Mpc}$ (green); the legend reports the number of systems in each bin. The grey dashed line marks the isotropic expectation ($\langle\theta\rangle=45^\circ$). Points show the binned mean values of the angle $\theta$ between the satellite orbital-normal vector $\hat{\boldsymbol{n}}_{\rm orb}$ and the projected filament direction $\hat{\boldsymbol{e}}_{\rm fil}$, and error bars indicate the $1\sigma$ uncertainty on the mean estimated from bootstrap resampling.}
\label{fig:determin}
\end{figure*}

\subsection{Identifying the primary driver(s)}

We found that, in Figure~\ref{fig:signal_dependence},  the values of $\langle\theta\rangle < 45^{\circ}$ preferentially appear in systems with lower host mass, at larger group-centric radii, and in regions further from the filament spine. This raises the question of which of these properties is the primary driver. From a dynamical perspective, satellites at larger $r/R_{200}$ are less influenced by the host halo potential, and their orbits may be more susceptible to the impact of the larger-scale environment, such as the cosmic filament. To disentangle these competing effects, Figure~\ref{fig:determin} shows the mean alignment angle $\langle\theta\rangle$ as a function of group richness $N_{\rm gal}$ (left panel),  normalized radial distance $r/R_{200}$ (middle panel), and host halo mass $M_{200}$ (right panel),  which similar to Figure~\ref{fig:signal_dependence} but with the sample subdivided into three bins of filament distance: $D_{\rm fil} < 0.5\,h^{-1}\mathrm{Mpc}$ (blue), $0.5 \le D_{\rm fil} < 1.0\,h^{-1}\mathrm{Mpc}$ (orange), and $1.0 \le D_{\rm fil} < 10.0\,h^{-1}\mathrm{Mpc}$ (green). This controlled comparison allows us to investigate whether the dependencies identified in the full sample are intrinsic or are secondary effects driven by the distance to the filament.

The left panel of Figure~\ref{fig:determin} illustrates the mean alignment angle $\langle\theta\rangle$ as a function of group richness $N_{\rm gal}$, split by filament distance. The blue curve ($D_{\rm fil} < 0.5\,h^{-1}\mathrm{Mpc}$) is consistently located above the isotropic expectation of $45^\circ$ across most of the richness range. It starts at $\langle\theta\rangle \approx 47^\circ$ for group richness $N_{\rm gal} = 2$, and increases to $\sim 49^\circ - 51^\circ$ for $N_{\rm gal} \approx 5-10$. It is worth noting that for systems with small $N_{\rm gal}$, the determination of the satellite orbital plane is less robust due to the limited number of satellites; nonetheless, a clear alignment signal persists even in these relatively poor groups as long as they are close to the filament spine. In contrast, the orange curve ($0.5 \le D_{\rm fil} < 1.0\,h^{-1}\mathrm{Mpc}$) shows a markedly different behavior, generally staying near or below the isotropic line. For the poorest groups at this intermediate distance, the signal drops significantly to $\sim 39^\circ$, suggesting orbital planes that are preferentially perpendicular to the filament, before fluctuating toward $45^\circ$ as richness increases. Finally, the green curve ($1.0 \le D_{\rm fil} < 10.0\,h^{-1}\mathrm{Mpc}$) remains relatively flat and close to the isotropic expectation across the entire richness range, with only small variations. These results suggest that the overall increase of $\langle\theta\rangle$ with richness seen in Figure~\ref{fig:signal_dependence} is primarily driven by groups located within the immediate vicinity of cosmic filaments.

In the middle panel of Figure~\ref{fig:determin}, we examine the radial dependence of the alignment signal across different filament distance bins. 
For the near-filament subsample (blue curve), $\langle\theta\rangle$ shows a non-monotonic trend, rising to a peak of $\sim 50^\circ$ around $r/R_{200} \approx 1.5$ before declining at larger radii, while remaining above the isotropic line across the radial range shown. This indicates that the positive alignment signal is strongest for systems located closest to the filament spine, although it is partially reduced at both small and large group-centric radii. For the intermediate-distance subsample (orange curve), $\langle\theta\rangle$ declines overall with increasing $r/R_{200}$, dropping from values near the isotropic expectation at small radii to a minimum of $\langle\theta\rangle \approx 38^\circ$ at $r/R_{200} \approx 3$, and remaining well below $45^\circ$ thereafter. This indicates that, at intermediate distances from the filament spine, satellite orbital planes at large group-centric radii are preferentially oriented perpendicular to the filament axis. Finally, the far-filament subsample (green curve) remains close to the isotropic value of $45^\circ$ over the full radial range, with only a weak declining trend at larger radii. The weak signal in this subsample suggests that the filament-related kinematic coherence becomes substantially weaker for systems located farther from the filament spine.

By contrast, both the orange and green curves lie close to the isotropic expectation ($\langle\theta\rangle\approx45^\circ$) in the innermost bins (small $r/R_{200}$). This behaviour is consistent with the interpretation that anisotropic accretion imposed by large-scale structure can be substantially washed out by nonlinear processes and dynamical mixing inside the host halo: satellites that have reached the inner regions show more randomized orbital-normal orientations regardless of their filament distance. The blue curve—representing satellites very near the filament spine—also shows a reduction of $\langle\theta\rangle$ toward smaller radii (relative to its peak), indicating that halo internal processes partially suppress the large-scale alignment even for these systems; however, the suppression is incomplete and the near-filament subsample retains a net $\langle\theta\rangle>45^\circ$ over much of the inner-to-intermediate radial range. Together, these patterns suggest that while internal halo dynamics tend to erase externally imprinted anisotropies, residual filament-driven coherence persists most strongly for satellites that remain close to the filament spine.

The right panel of Figure~\ref{fig:determin} presents the dependence of the alignment signal on the host halo mass $M_{200}$. For the near-filament subsample (blue curve), we find a robust and monotonic increase in $\langle\theta\rangle$ with host mass. At the low-mass end ($M_{200} \lesssim 10^{11.5}\,M_\odot/h$), the signal is only slightly above the isotropic expectation, but it rises steadily to reach $\langle\theta\rangle \approx 52^\circ$ for the most massive clusters ($M_{200} \gtrsim 10^{14}\,M_\odot/h$). This suggests that more massive host haloes, with their deeper gravitational potentials and larger cross-sections for accretion, are more effective at channeling satellites along their parent filaments into orbits that preferentially contain the filament axis. 

In sharp contrast, for the intermediate-distance subsample (orange curve), the mean alignment angle remains consistently below $45^\circ$ for nearly the entire mass range. Although there is a subtle upward trend in the highest mass bins, the signal fails to recover to the isotropic value, peaking at only $\sim 44^\circ$. This indicates that for satellites at intermediate distances ($0.5 \le D_{\rm fil} < 1.0\,h^{-1}\mathrm{Mpc}$), even a significant increase in host mass is insufficient to overcome the environment-driven preference for orbital planes that are perpendicular to the filament. The far-filament subsample (green curve) remains remarkably stable near $44^\circ - 45^\circ$, showing almost no mass dependence. These results confirm that while host halo mass modulates the strength of the alignment, its impact is secondary to that of filament distance; a strong positive alignment signal ($\langle\theta\rangle > 45^\circ$) is only realized when satellites are in close distance to the filament spine, with the magnitude of this effect being amplified by the mass of the host.

Overall, the controlled comparisons in Figure~\ref{fig:determin} demonstrate that $D_{\rm fil}$ is the primary determinant of the kinematic alignment signal's sign and fundamental orientation. While richness, radial distance, and halo mass all modulate the signal’s amplitude, their effects are largely conditional on the satellite's location relative to the cosmic filament, supporting a scenario where filamentary accretion dictates the initial orbital configuration which is subsequently modified by host halo properties and internal dynamics.

\section{Summary}\label{sec:sum}
To investigate the modes of matter accretion onto filaments and their subsequent transfer to galaxy groups/clusters, specifically to observationally test the two-phase model \citep[][]{2015ApJ...813....6K,2018MNRAS.473.1562W} proposed by simulation studies and supported by emerging circumstantial evidence from observations \citep{2026ApJ...999..158Y}, we use SDSS DR12 galaxies, the Bisous filament catalogue, and a Friends-of-Friends (FoF) group catalogue to measure the relative orientation between the projected satellite orbital-normal vector and the projected filament direction for a fiducial sample of 76,959 galaxy systems.

We summarize the main findings of this work as follows:
\begin{enumerate}
\item The distribution of the relative orientation angle $\theta$ shows a statistically significant excess at large angles compared to an isotropic expectation. The mean value for fiducial sample is $\langle\theta\rangle=46.15^\circ\pm0.09^\circ$, inconsistent with isotropy at $\sim12.8\sigma$ (KS $p\sim4.6\times10^{-33}$), indicating that, overall, the accretion of satellite galaxies is not isotropic but is instead channeled through the cosmic filament.
\item Filament distance ($D_{\rm fil}$) is the primary driver of this trend: starting from isotropy at large separations, the mean angle first drops to $\langle\theta\rangle \lesssim 41.5^\circ$ at intermediate distances ($0.5 \lesssim D_{\rm fil} \lesssim 1 \, h^{-1}{\rm Mpc}$), before sharply increasing to a maximum of $\langle\theta\rangle \sim 56^\circ$ for systems very close to the filament spine ($D_{\rm fil} \lesssim 0.1 \, h^{-1}{\rm Mpc}$), implying a transition from transverse infall onto filaments to longitudinal channeling along them.
\item Host halo mass and group richness modulate the signal: more massive and richer systems exhibit stronger alignment with filaments. The radial dependence is non-monotonic: the alignment peaks at intermediate radii ($r\sim1.5$-$2\,R_{200}$), weakens in the most inner radii, and weakens or reverses in the outskirts ($r\gtrsim5\,R_{200}$).
\end{enumerate}

These results indicate that cosmic filaments leave a measurable kinematic imprint on satellite systems.  Those satellite systems in the outskirts of low-mass hosts, and those located relatively far from a filament's spine tend to be accreted along directions more nearly perpendicular to the filament.  Moving inward, however, satellites accreted along filamentary channels preferentially orbit in planes that contain the filament spine, reflecting the anisotropic nature of mass assembly.  This transition reveals a complex dynamical process supporting a two-phase accretion picture: in the first phase, mass is captured from the surrounding environment via transverse infall towards the filament spine; subsequently, the second stage manifests as galaxies being coherently channeled along the filamentary corridors into the clusters.

\section{Discussion}\label{sec:dis}

Our detection of a statistically significant correlation between satellite orbital configurations and the local filament direction---characterized by a flip in orientation preference from perpendicular to parallel as a function of filament distance---provides a complementary kinematic perspective on how cosmic filaments influence galaxy assembly. To our knowledge, this work presents the first observational detection of a transition in the accretion geometry around filaments. This transition reveals a complex dynamical process supporting a two-phase accretion picture \citep[][]{2018MNRAS.473.1562W}. 

The scale-dependent behaviour of $\langle\theta\rangle$ we report is consistent with the two-phase accretion around filaments. At large separations ($D_{\rm fil}\sim10\ h^{-1}\mathrm{Mpc}$) the signal is close to isotropic, indicating a weak filamentary influence on orbital geometry.  At intermediate distances the mean angle falls below $45^\circ$, reaching a minimum around $D_{\rm fil}\sim0.5$--$0.7\ h^{-1}\mathrm{Mpc}$; this regime may correspond to a characteristic filament radius \citep[By adopting the same methodology as][, we got radius of $0.56 \ \rm h^{-1}\mathrm{Mpc}$ for our filament sample]{2024MNRAS.532.4604W} where transverse infall from surrounding cosmic walls or sheets and interactions between nearby branches impart orbital angular-momentum components that favour orbital planes perpendicular to the filament axis.  Deeper inside the filamentary spine ($D_{\rm fil}\lesssim0.3\ h^{-1}\mathrm{Mpc}$) the signal flips to $\langle\theta\rangle>45^\circ$, consistent with satellites becoming more tightly coupled to coherent, along-filament streaming and to the filament's internal circulation.  Put another way, the measured ``swing'' of $\langle\theta\rangle$ around the isotropic expectation ($45^\circ$) with $D_{\rm fil}$ maps naturally onto a physical sequence in which satellites transition from near-isotropic outer regions, through a zone dominated by transverse accretion and reorientation, into a regime dominated by coherent, filament-aligned streaming and tangential circulation.

In this two-phase accretion scenario, namely ``helical -like'' mass flow around filament, the interplay between tangential motions around and streaming motions along the filament axis has distinct dynamical roles: the tangential component of satellite velocities about the filament spine directly feeds the filament's angular momentum as a whole \citep[i.e.\ its spin by][]{2021NatAs...5..839W,2025ApJ...982..197T,2025ApJ...983..100W}, whereas longitudinal streaming along the spine supplies a directed flux \citep{2026arXiv260118434Y} of mass and momentum into the host group or cluster. Moreover, the combined helical  trajectories naturally produce a evolution of the galaxy spin--filament alignment: as satellites are captured and increasingly subjected to both tangential shear and longitudinal streaming when approaching the spine, their spin orientations are reoriented, yielding the observed flip \citep{2014MNRAS.444.1453D,2018MNRAS.473.1562W} in the galaxy spin--filament correlation (from perpendicular to parallel) without the need to invoke additional, separate physical mechanisms.

These observational findings complement and extend prior simulation-based studies of galaxy spin–filament transitions and filament-driven dynamics \citep[e.g.,][]{2015ApJ...813....6K,2018MNRAS.473.1562W,2014MNRAS.441.2923C}. Unlike measurements of halo or central-galaxy spin, the orbital angular momentum of satellite ensembles is a direct dynamical tracer of anisotropic accretion, and thus provides an independent line of evidence that the cosmic web imprints kinematic coherence across a range of scales. The stronger signal near filament spines and in more massive, higher-richness hosts aligns with expectations that (i) massive haloes have larger capture cross sections and deeper potentials that preserve coherence of filament-fed satellites, and (ii) richer groups supply a larger satellite sample that stabilizes the orbital-normal estimate. The interpretation that filaments both spin \citep[via tangential motions:][]{2021NatAs...5..839W} and feed \citep[via along-filament streaming:][]{2014MNRAS.443.1274L,2026arXiv260118434Y} is furthermore attractive because it links the local kinematic patterns we observe to global angular-momentum and mass-transport processes operating within the cosmic web.

We emphasize several observational caveats. Our estimator relies on projected positions and line-of-sight velocities, so projection effects and redshift-space distortions (RSDs) can bias filament assignment and the inferred orbital-normal vectors; FoF group-finding and central identification uncertainties, interloper contamination, and finite-satellite sampling (small $N_{\rm gal}$) all affect the noise and potential bias of the measurement. The choice of filament-finder \footnote{We use the Bisous, we refer readers to \cite{2018MNRAS.473.1195L} for a discussion of various filament finders} can imprint algorithm-dependent spine orientations and distances (see Appendix~\ref{app:sec1} for a comparative testing with DisPerSE), and $M_{200}$ estimates derived from velocity dispersions carry scatter and bias that influence mass-dependent trends. We have performed randomized-shuffle null tests and bootstrap resampling to check for simple chance alignments, but a full quantification of systematics requires realistic mock catalogues that fold in survey selection, RSDs, group-finder performance and the filament detection pipeline.

Follow-up work will leverage large N-body and hydrodynamical simulations for two primary purposes. First, synthetic mock catalogues will be generated and processed through identical pipelines to quantify projection and selection biases, ensuring the robustness of current trends. Second, direct 3D diagnostic analyses will trace satellite orbits and decompose velocity components relative to filaments. These simulation-based efforts are essential to distinguish between tangential circulation and longitudinal streaming, thereby identifying the physical mechanisms driving the observed orientation transitions. Future extensions to high-redshift surveys  and the inclusion of stellar populations or proper-motion data will further clarify the interplay between cosmic-web dynamics and galaxy evolution.

\begin{acknowledgments}
P.W. acknowledges the financial support from the NSFC (No. 12595312, No. 12473009), and is also sponsored by Shanghai Rising-Star Program (No. 24QA2711100). This work is supported by the China Manned Space Program with grant no. CMS-CSST-2025-A03. 
M.B. acknowledges support by the National Natural Science Foundation of China, NSFC grant No. 12303009.
YC.Z. acknowledges the financial support from the NSFC (No. 12273088).
\end{acknowledgments}

\appendix
\setcounter{figure}{0}
\renewcommand{\thefigure}{A\arabic{figure}} 

\section{Test using an alternative filament finder}
\label{app:sec1}

\begin{figure}[t]
\centering
\plotone{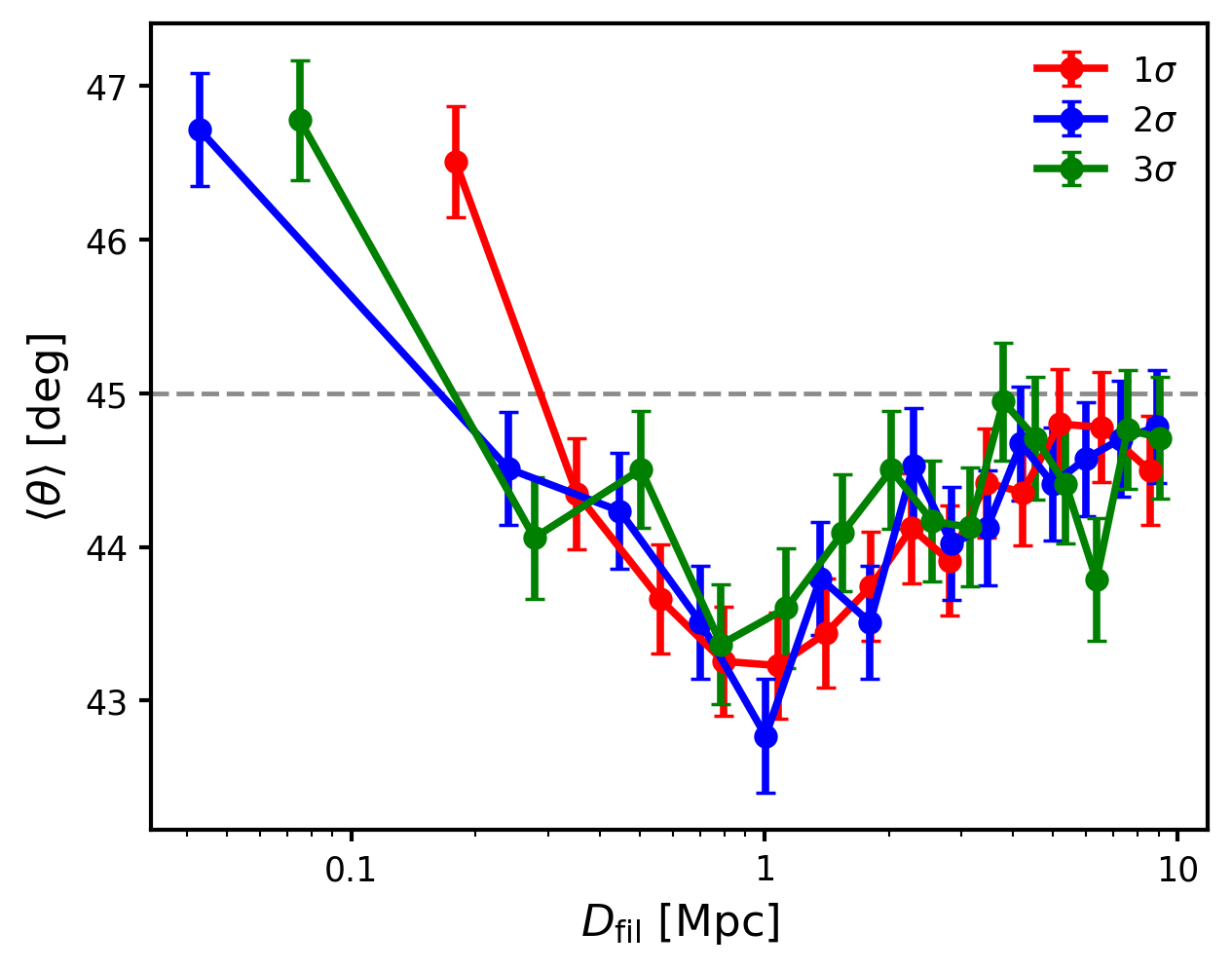}
\caption{Mean angle \(\langle\theta\rangle\) as a function of distance to the nearest filament spine \(D_{\rm fil}\), measured using filaments identified with DisPerSE at three persistence thresholds. Red/blue/green curves show results for persistence levels labelled 1\(\sigma\), 2\(\sigma\), and 3\(\sigma\), respectively; all filament catalogs are constructed from the same galaxy sample. Error bars indicate measurement uncertainties. The horizontal dashed line marks \(\langle\theta\rangle=45^\circ\) (random orientations).
}
\label{fig:figA1}
\end{figure}

In this section, we examine the robustness of our results with respect to the chosen filament identification algorithm. Since different finders employ distinct mathematical frameworks to trace the cosmic web, we repeat our analysis using DisPerSE \citep[][]{2011MNRAS.414..350S,2011MNRAS.414..384S} with three persistence thresholds ($1\sigma$, $2\sigma$, and $3\sigma$). As shown in Figure~\ref{fig:figA1}, the qualitative trend remains consistent across all persistence levels: a notable alignment signal is recovered at intermediate scales ($0.3 \lesssim D_{\text{fil}} \lesssim 1\,\text{Mpc}$), confirming that the detected signal is a robust feature that does not depend on the specific topological sensitivity of the finder.

A key consideration in this test is the use of the full galaxy sample (including satellites) for filament reconstruction. It is important to recognize that the distribution of satellite galaxies and the definition of filamentary structures are inherently coupled at a physical level. Satellites are not merely external tracers; they are dynamical components of the large-scale structure that tend to reside within and stream along the filamentary spines. Consequently, including these galaxies in the filament-finding process naturally reveals a more intricate and detailed network compared to using sparse tracers.

This coupling is an intrinsic property of the cosmic web. Excluding satellites to ``de-bias'' the filament recovery would be conceptually similar to excluding halo particles when defining the large-scale gravitational environment (e.g., via the T-web); such an approach would omit the very matter that constitutes the structure. The variation in alignment amplitude across different persistence levels reflects this multi-scale nature: lower thresholds capture the smaller branches traced by local galaxy clustering, while higher thresholds focus on the high-significance skeletons. The stability of the overall trend suggests that while the fine details of the filament spines shift depending on the tracer density and algorithm parameters, the underlying physical alignment between galaxies and their host environment remains a prominent and recoverable signal.

\bibliography{main}{}
\bibliographystyle{aasjournal}


\end{CJK*}
\end{document}